\documentclass{emulateapj}

\def \deg{\hbox{$^\circ$}}

\shorttitle{Insights on the Torus from a Complete Volume Limited Hard X-ray AGN Sample}
\shortauthors{Davies et al.}

\begin{document}


\title{Insights on the Dusty Torus and Neutral Torus from Optical and X-ray Obscuration in a Complete Volume Limited Hard X-ray AGN Sample}


\author{
R.I.~Davies\altaffilmark{1},
L.~Burtscher\altaffilmark{1},
D.~Rosario\altaffilmark{1},
T.~Storchi-Bergmann\altaffilmark{2},
A.~Contursi\altaffilmark{1},
R.~Genzel\altaffilmark{1},
J.~Graci\'a-Carpio\altaffilmark{1},
E.~Hicks\altaffilmark{3},
A.~Janssen\altaffilmark{1},
M.~Koss\altaffilmark{4},
M.-Y.~Lin\altaffilmark{1},
D.~Lutz\altaffilmark{1},
W.~Maciejewski\altaffilmark{5},
F.~M\"uller-S\'anchez\altaffilmark{6},
G.~Orban~de~Xivry\altaffilmark{1},
C.~Ricci\altaffilmark{7,8},
R.~Riffel\altaffilmark{2},
R.A.~Riffel\altaffilmark{9},
M.~Schartmann\altaffilmark{10},
A.~Schnorr-M\"uller\altaffilmark{1},
A.~Sternberg\altaffilmark{11},
E.~Sturm\altaffilmark{1},
L.~Tacconi\altaffilmark{1},
S.~Veilleux\altaffilmark{12}
}

\altaffiltext{1}{Max-Planck-Institut f\"ur extraterrestrische Physik, Postfach 1312, 85741, Garching, Germany}
\altaffiltext{2}{Departamento de Astronomia, Universidade Federal do Rio Grande do Sul, IF, CP 15051, 91501-970 Porto Alegre, RS, Brazil}
\altaffiltext{3}{Astronomy Department, University of Alaska, Anchorage, USA}
\altaffiltext{4}{Institute for Astronomy, Department of Physics, ETH Zurich, Wolfgang-Pauli-Strasse 27, CH-8093 Zurich, Switzerland}
\altaffiltext{5}{Astrophysics Research Institute, Liverpool John Moores University, IC2 Liverpool Science Park, 146 Brownlow Hill, L3 5RF, UK}
\altaffiltext{6}{Center for Astrophysics and Space Astronomy, University of Colorado, Boulder, CO 80309-0389, USA}
\altaffiltext{7}{Pontificia Universidad Cat\'olica de Chile, Instituto de Astrof\'isica, Casilla 306, Santiago 22, Chile}
\altaffiltext{8}{EMBIGGEN Anillo, Concepci\'on, Chile}
\altaffiltext{9}{Departamento de F\'isica, Centro de Ci\^encias Naturais e Exatas, Universidade Federal de Santa Maria, 97105-900 Santa Maria, RS, Brazil}
\altaffiltext{10}{Centre for Astrophysics and Supercomputing, Swinburne University of Technology, Hawthorn, Victoria, 3122, Australia}
\altaffiltext{11}{Raymond and Beverly Sackler School of Physics \& Astronomy, Tel Aviv University, Ramat Aviv 69978, Israel}
\altaffiltext{12}{Department of Astronomy and Joint Space-Science Institute, University of Maryland, College Park, MD 20742-2421 USA}

\begin{abstract}
We describe a complete volume limited sample of nearby active galaxies selected by their 14--195\,keV luminosity, and outline its rationale for studying the mechanisms regulating gas inflow and outflow.
We describe also a complementary sample of inactive galaxies, selected to match the AGN host galaxy properties.
The active sample appears to have no bias in terms of AGN type, the only difference being the neutral absorbing column which is two orders of magnitude greater for the Seyfert\,2s.
In the luminosity range spanned by the sample, $\log{L_{14-195\,keV} [erg\,s^{-1}]} = 42.4$--43.7, the optically obscured and X-ray absorbed fractions are 50--65\%.
The similarity of these fractions to more distant spectroscopic AGN samples, although over a limited luminosity range, suggests that the torus does not strongly evolve with redshift.
Our sample confirms that X-ray unabsorbed Seyfert~2s are rare, comprising not more than a few percent of the Seyfert~2 population.
At higher luminosities, the optically obscured fraction decreases (as expected for the increasing dust sublimation radius), but the X-ray absorbed fraction changes little.
We argue that the cold X-ray absorption in these Seyfert~1s can be accounted for by neutral gas in clouds that also contribute to the broad line region (BLR) emission;
and suggest that a geometrically thick neutral gas torus co-exists with the BLR and bridges the gap to the dusty torus.
\end{abstract}


\section{Introduction}
\label{sec:intro}

Feeding and feedback have become a paradigm of galaxy evolution models: by quenching star formation, feedback from active galactic nuclei (AGN) is thought to shape the galaxy luminosity function and create the bi-modal colour sequence in galaxy
populations \citep{ben03,kau03}.
But the prescriptions used in models \citep{spr05,cro06,som08} are relatively simple because observations have focussed on the question of where the inflowing material originates (mergers versus secular evolution) and on integrated galaxy properties. 
To redress this, we need, in addition to large scale data, also to understand how the material flows towards the black hole (BH) on smaller scales. 
However these scales cannot be spatially resolved at $z > 1$ where co-evolution largely occurs \citep{fab12}.
Instead, it is local galaxies that currently offer the only opportunity to guide the small scale aspects of galaxy and BH co-evolution models.
That many local AGN -- in particular Seyferts -- are disky systems does not necessarily reduce their validity as templates for more distant galaxies.
The nested simulations of \cite{hop10} provide an important insight in this respect: they suggest that as one looks further inside the central kiloparsec, disk processes become increasingly important in driving gas inwards,
independent of what has occurred on large scales. 
Observations also appear to confirm that even at $z=1$--2 disk processes are a major contributor to AGN fueling \citep{koc12,sch12,kar14,vil14}.

In nearby archetypal objects, integral field spectroscopy (at optical and infrared wavelengths, and sometimes with adaptive optics) has probed kinematics on 10--100\,pc scales, leading to insights in both inflow (see \citealt{sto14a}) and outflow (see \citealt{sto14b}).
Despite the inherent complexities of performing detailed systematic studies with such techniques, and because of the large amount of observing time needed, there has been some progress using integral field spectroscopy for very small samples of AGN \citep{sos01,bar06,dav07,hic09,bar09,mue11,rup11,rup13,mue15}, and in a few cases with a matched sample of inactive galaxies for comparison \citep{dum07,wes12,hic13,dav14}.
There are also two larger studies of local AGN based on integral field spectroscopy, but which do not address spatially resolved structures or kinematics: \cite{bur15} focus on the spatially unresolved near-infrared non-stellar continuum in Seyferts and other low luminosity AGN (using the spatial information only to help characterise this component); \cite{stu11} and \cite{vei13} focus on ULIRGs and QSOs, for which the data are spatially unresolved due to the long wavelengths observed.

In this paper, we present a complete, volume limited sample of nearby bright hard X-ray selected AGN, in a luminosity range that overlaps with AGN at higher redshift.
This statistically meaningful sample of 20 AGN is complemented by a sample of inactive galaxies that are matched in mass, morphology and inclination.
In Section~\ref{sec:sample} we describe the scientific rationale for, and the selection of, both the active and inactive samples; 
and outline the observations that have been started in order to spatially and spectrally resolve the processes that drive and regulate gas inflow and outflow on small scales.
We then look at the AGN properties in Section~\ref{sec:agntype}, in particular the fraction of Sy\,1s and Sy\,2s, and the neutral gas columns.
Reconciling this with the numbers reported in the literature leads to a discussion of optical obscuration and X-ray absorption.
Section~\ref{sec:obsc} focusses on the issue of X-ray unabsorbed Sy\,2s.
Then in Section~\ref{sec:torus} we look at what the differing luminosity dependencies of the optically obscured and X-ray absorbed fractions can tell us about the torus and broad line region (BLR).
We have chosen this approach -- avoiding the need to invoke torus models -- because current torus models are in need of additional basic constraints.
As pointed out by \cite{hoe13} and \cite{fel12}, different models can come to contradictory conclusions and model parameters are often degenerate.

\begin{figure*}
\epsscale{0.90}
\plotone{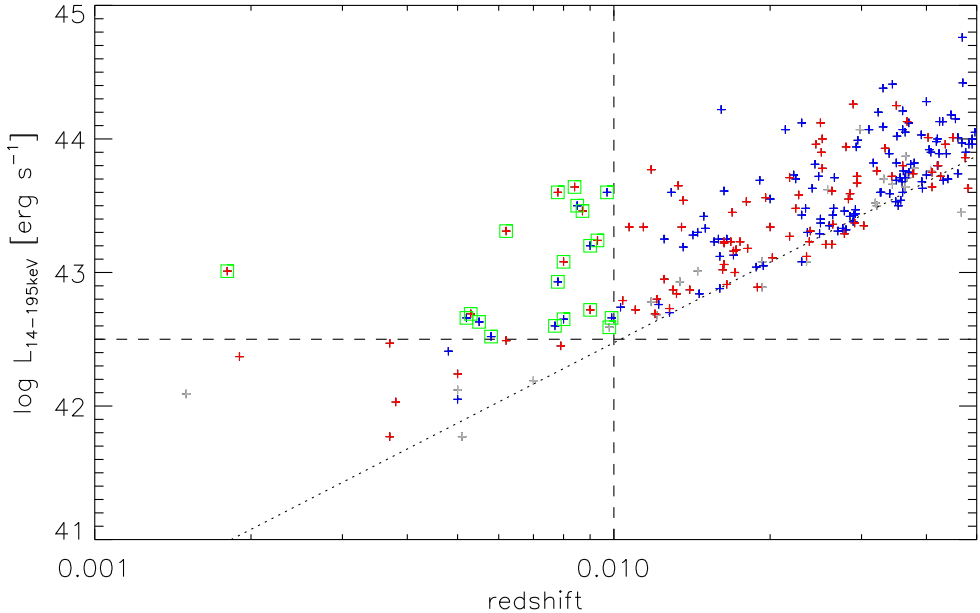}
\caption{\label{fig:batselect}
Plot of the redshift versus luminosity for AGN in the 58-month {\it Swift-BAT} survey that are readily observable from the VLT, i.e. with $\delta < 15\deg$, truncated at $z=0.05$.
These have been colour coded blue/red for Sy\,1/Sy\,2 where such classifications are available (note that only `class 4' AGN are within our luminosity and redshift range).
The dotted line is the flux limit for 90\% of the sky.
The dashed lines indicate the distance and luminosity thresholds for our selection in order to have a complete sample -- for which the selected targets are marked with green.
Luminosities in this plot are based on distances derived from the uncorrected redshift, as used in the initial target selection. These may differ slightly from the distances listed in Table~\ref{tab:agn} which are, where possible, redshift independent or corrected for peculiar velocities.}
\end{figure*}

\section{Sample}
\label{sec:sample}

\renewcommand{\thefootnote}{\alph{footnote}}

Most of the studies in the literature that focus on the structure and processes occurring in the circumnuclear regions of samples of local AGN, have selected objects using the optical classification (e.g. as Seyfert or LINER) as the only defining characteristic of the AGN.
An obvious reason for doing so is that there is a wide variety of local AGN and such studies often try to span a range in AGN types.
However, this may lead to confusing results because of the impact of accretion rate, or equivalently and as a proxy for it, luminosity \citep{ho08,dav12,kor13}.
For example, a bright Seyfert such as NGC\,1068 with 
$\log{L_{AGN}} \sim 45$\footnote[1]{Throughout this paper we use units of erg\,s$^{-1}$ for luminosities.
$L_{AGN}$ refers to the bolometric luminosity of an AGN.
$L_{14-195}$ is the observed luminosity in the 14--195\,keV band.
$L_{2-10}$ and $L^{int}_{2-10}$ are respectively the observed and absorption corrected (i.e. intrinsic) luminosities in the 2--10\,keV band.
$L_{12\mu m}$ refers to the monochromatic nuclear 12$\mu$m luminosity, measured at subarcsecond spatial resolution (for details see \citealt{asm14}).
We also adopt the relations 
$\log{L_{AGN}} = 1.12 \log{L_{14-195}} - 4.23$ \citep{win12} and 
$\log{L^{int}_{2-10}}=1.06\log{L_{14-195}}-3.08$ \citep{win09}.} requires an accretion rate of order $0.1\,M_\odot$\,yr$^{-1}$, while a low luminosity AGN such as M\,81 with $\log{L_{AGN}} \sim 41$ will have an accretion rate $\sim10^{-5}\,M_\odot$\,yr$^{-1}$ that is 4 orders of magnitude lower.
This difference has a dramatic impact on fuelling requirements.
\cite{hic09} and \cite{maz13} show that one can expect to find a gas mass of order $10^7$\,M$_\odot$ within a radial scale of $\sim10$\,pc.
Assuming a 1\% efficiency to bring this to accretion disk scales \citep{mue09,sch10}, this mass alone is sufficient to fuel M\,81 for 10\,Gyr, but it would supply NGC\,1068 for only 1\,Myr.
In this context, an accretion rate similar to that for NGC\,1068 is clearly very significant and may only be sustained via a relatively efficient inflow, perhaps driven by a coherent dynamical mechanism on scales of 0.1-1\,kpc where the gas reservoir is $10^8$--$10^9$\,M$_\odot$ \citep{sak99,maz13}.
In contrast, low accretion rates comparable to that for M\,81 could be supplied either by a gas reservoir on small scales, or by rather inefficient inflow from circumnuclear scales.
Observationally, such an effect is implicit in the relation between star formation age and accretion rate presented by \cite{dav07}: while AGN with higher luminosities ($\sim10^{45}$\,erg\,s$^{-1}$) or Eddington ratios (0.1--1) may be associated with young post-starbursts, this is not the case for AGN with luminosities two orders of magnitude lower.
It could also help to explain the results of a survey of molecular gas \citep{haa09,gar11} which found that gravitational torques could drive gas inwards for not more than half of their sample.
And the role of AGN luminosity has been highlighted in the context of molecular outflows, in the sense that only very luminous AGn drive massive outflows \citep{stu11,vei13}.

Our premise for defining a new sample is that the black hole accretion rate, traced by the AGN luminosity, is a key issue that needs to be considered in order to provide a context both for the mechanisms driving and regulating inflow and outflow as well as the derived flow rates.
To address this issue, our sample is based on bright local AGN in which the accretion rate is relatively high.
We plan to analyse them in the context of matching inactive galaxies, for which the accretion rate is, by definition, orders of magnitude lower.

An obvious concern here is the timescale of AGN variability \citep{nov11,hic14} with respect to the timescale of the phenomenon (e.g. starburst, dynamical process, etc) that is being assessed in the context of accretion onto the black hole.
If an inactive sample is used naively as a control in a direct comparison to the active sample, the implicit assumption must be that any link to accretion is strong enough to persist even when averaged over the lifetime of the phenomenon being studied.
However, \cite{dav14} argue and show that with a more judicious use of an inactive sample, one can make progress even if this assumption is not strictly met.

Building on the smaller sample of \cite{hic13} and \cite{dav14}, and also on the work summarised by \cite{sto14a,sto14b}, we plan to address a few key questions in a statistically robust way:\smallskip

1) Does star formation play a decisive role in either driving or hindering gas inflow to AGN? This question focusses on the central 100\,pc, assessing whether there has been recent star formation, and if so whether it is still on-going or has ceased.\smallskip

2) What mechanisms are responsible for driving gas from the kiloparsec scale into the central tens of parsecs, and what triggers these? This includes dynamical processes in the central kiloparsec as well as the role of the larger scale host galaxy and the influence of the local environment.\smallskip

3) Do luminous Seyferts always drive ionised and/or molecular outflows, and how do they interact with the interstellar medium? This addresses the ubiquity and efficacy of a variety of emission lines as tracers of outflows, as well as quantifying the outflow velocities, rates, and mass loading.\smallskip

These -- and other -- questions will be addressed in future papers using a combination of high spectral resolution data from 0.4--2.3$\mu$m taken with XShooter \citep{ver11} and high spatial resolution integral field spectroscopy in the H- and K-bands taken with SINFONI \citep{eis03,bon04}, both on the VLT.
At the time of writing, XShooter observations have been completed for 8 Seyferts, and 10 inactive galaxies, and the remainder of the targets are scheduled; SINFONI observations are completed or scheduled for the first half of the sample.

\begin{deluxetable*}{lclccccclcc}
\tabletypesize{\scriptsize}
\tablecaption{Summary of AGN and host galaxy properties\label{tab:agn}} 
\tablewidth{0pt}
\tablehead{
\colhead{Name} &
\colhead{Distance} &
\colhead{AGN} &
\colhead{log L$_{H}$} &
\colhead{log L$_{14-195}$} &
\colhead{log L$_{2-10}$} &
\colhead{log L$_{12\mu m}$} &
\colhead{log N$_H$} &
\colhead{Hubble} &
\colhead{axis} &
\colhead{bar} \\
\colhead{} &
\colhead{(Mpc)} &
\colhead{classification} &
\colhead{{L$_\odot$}} &
\colhead{erg\,s$^{-1}$} &
\colhead{erg\,s$^{-1}$} &
\colhead{erg\,s$^{-1}$} &
\colhead{cm$^{-2}$} &
\colhead{stage} &
\colhead{ratio} &
\colhead{} \\
\colhead{(1)} &
\colhead{(2)} &
\colhead{(3)} &
\colhead{(4)} &
\colhead{(5)} &
\colhead{(6)} &
\colhead{(7)} &
\colhead{(8)} &
\colhead{(9)} &
\colhead{(10)} &
\colhead{(11)} \\
}
\startdata

NGC 1365      &	18  & Sy 1.8\tablenotemark{a}    &    10.58 & 42.39 & 41.83 & 42.54 & \phm{$\leq$}22.2\tablenotemark{d} & \phm{\ \ -}3   & 0.55 &  B \\
MCG-05-14-012 &	41  & Sy 1.0\tablenotemark{a}    & \phn9.60 & 42.60 & 41.63 &       & $\leq$21.9\phm{$^a$}              & \phm{\ \ }-1   & 0.86 &    \\
NGC 2110      &	34  & Sy 2 (1h)\tablenotemark{a} &    10.44 & 43.64 & 42.53 & 43.04 & \phm{$\leq$}23.0\tablenotemark{e} & \phm{\ \ }-3   & 0.74 & AB \\
NGC 2992      &	36  & Sy 1.8\tablenotemark{a}    &    10.31 & 42.62 & 42.05 & 42.87 & \phm{$\leq$}21.7                  & \phm{\ \ -}1   & 0.30 &    \\
MCG-05-23-016 &	35  & Sy 1.9\tablenotemark{a}    & \phn9.94 & 43.47 & 43.11 & 43.42 & \phm{$\leq$}22.2\phm{$^a$}        & \phm{\ \ }-1   & 0.45 &    \\
NGC 3081      &	34  & Sy 2 (1h)\tablenotemark{a} &    10.15 & 43.06 & 41.54 & 42.75 & \phm{$\leq$}23.9\phm{$^a$}        & \phm{\ \ -}0   & 0.77 & AB \\
NGC 3783      &	38  & Sy 1.2\tablenotemark{a}    &    10.29 & 43.49 & 43.12 & 43.47 & \phm{$\leq$}20.5\phm{$^a$}        & \phm{\ \ -}1.5 & 0.89 &  B \\
NGC 4235      &	37  & Sy 1.2                     &    10.43 & 42.72 & 41.66 & 42.17 & \phm{$\leq$}21.3\phm{$^a$}        & \phm{\ \ -}1   & 0.22 &    \\
NGC 4388      &	39  & Sy 2 (1h)                  &    10.65 & 43.70 & 42.57 & 42.93 & \phm{$\leq$}23.5\tablenotemark{d} & \phm{\ \ -}3   & 0.18 &  B \\
NGC 4593      & 37  & Sy 1.0--1.2\tablenotemark{a} &  10.59 & 43.16 & 42.77 & 42.97 & $\leq$19.2\phm{$^a$}              & \phm{\ \ -}3   & 0.74 &  B \\
NGC 5128 (Cen A) & 3.8 & Sy 2                    &    10.22 & 42.38 & 41.50 & 41.82 & \phm{$\leq$}23.1\tablenotemark{d} & \phm{\ \ }-2   & 0.78 &    \\
ESO 021-G004  &	39  & Sy\tablenotemark{b}        &    10.53 & 42.49 & 41.21 &       & \phm{$\leq$}23.8\phm{$^a$}        & \phm{\ \ }-0.4 & 0.45 &    \\
MCG-06-30-015 &	27  & Sy 1.2                     & \phn9.59 & 42.74 & 42.51 & 42.87 & \phm{$\leq$}20.9\phm{$^a$}        & \phm{\ \ }-5   & 0.60 &    \\
NGC 5506      &	27  & Sy 2 (1i)                  &    10.09 & 43.32 & 42.91 & 43.28 & \phm{$\leq$}22.4\phm{$^a$}        & \phm{\ \ -}1   & 0.23 &    \\
NGC 5728      &	39  & Sy 2                       &    10.56 & 43.21 & 41.41 & 42.35 & \phm{$\leq$}24.2\phm{$^a$}        & \phm{\ \ -}1   & 0.57 &  B \\
ESO 137-G034  &	35  & Sy 2                       &    10.44 & 42.62 & 40.86 &       & \phm{$\leq$}24.3\phm{$^a$}        & \phm{\ \ -}0   & 0.79 & AB \\
NGC 6814      &	23  & Sy 1.5                     &    10.31 & 42.69 & 42.17 & 42.18 & \phm{$\leq$}21.0\phm{$^a$}        & \phm{\ \ -}4   & 0.93 & AB \\
NGC 7172      &	37  & Sy 2 (1i)\tablenotemark{c} &    10.43 & 43.45 & 42.53 & 42.88 & \phm{$\leq$}22.9\phm{$^a$}        & \phm{\ \ -}1.4 & 0.56 &    \\
NGC 7213      &	25  & Sy 1                       &    10.62 & 42.50 & 41.95 & 42.58 & $\leq$20.4\phm{$^a$}              & \phm{\ \ -}1   & 0.90 &    \\
NGC 7582      &	22  & Sy 2 (1i)\tablenotemark{c} &    10.38 & 42.67 & 41.12 & 42.81 & \phm{$\leq$}24.2\tablenotemark{d} & \phm{\ \ -}2   & 0.42 &  B \\

\enddata

\tablenotetext{a}{Confirmed or updated based on our available XSHOOTER data (Schnorr-M\"uller et al. in prep.). Note that NGC\,4593 is ambiguous because it is close to the boundary between Sy\,1 and Sy\,1.2; MCG-05-23-016 has weak broad H$\alpha$; NGC\,2110 is changed from `Sy\,2 (1i)' to a simple `Sy\,2', because the polarised broad line has a double peaked profile which is not seen at near-infrared wavelengths.}

\tablenotetext{b}{ESO\,021-G004 has no optical classification, and our XSHOOTER data for this object have not yet been taken.}

\tablenotetext{c}{We have added a `1i' qualifier to the classifications of NGC\,7172 because \cite{sma12} show evidence for broad Pa$\alpha$ and Br$\gamma$, and NGC\,7582 because \cite{reu03} find broad Br$\gamma$.}

\tablenotetext{d}{Variations in N$_H$ between Compton thick and thin have been reported for NGC\,1365 \citep{ris09} and NGC\,7582 \citep{bia09}. Compton thin variations in N$_H$ have been reported for NGC\,4388 \citep{elv04} and NGC\,5128 \citep{bec11}.}

\tablenotetext{e}{The absorbing column for NGC\,2110 includes partial covering absorbers, and the value given here is weighted by covering factor. Details are given in \cite{ric15}; see also \cite{eva07}.}

\tablecomments{Columns: 
(1) common name;
(2) distances are taken from NED, using redshift independent estimates or peculiar velocity corrections derived by \cite{the07} where possible (except for NGC\,5128 where the distance is from \citealt{har10}), and so may differ slightly from the distance based on the uncorrected redshift used in the initial selection;
(3) AGN type as given in NED, except where indicated for the 8 AGN that we can independently classify, with additional information: 1i = near-infrared broad lines, 1h = polarised broad line emission;
(4) integrated H-band luminosity (given by the 2MASS total magnitude; \citealt{skr06}) as a proxy for stellar mass;
(5) observed 14--195\,keV luminosity (70\,month average) from {\it Swift-BAT} \citep{bau13};
(6) observed 2--10\,keV luminosity (single epoch) from \cite{ric15};
(7) nuclear 12\,$\mu$m luminosity from \cite{asm14} (MCG-05-14-012, ESO 021-G004, and ESO\,137-G034 are not included in this catalogue and, as far as we know, do not have 12\,$\mu$m measurements on arcsec scales);
(8) neutral absorbing column, from \cite{ric15} based on modelling 0.3-150\,keV spectrum;
(9) Hubble stage (NED);
(10) axis ratio (NED);
(11) B and AB denote presence of a bar or weak bar respectively (NED).}

\end{deluxetable*}

\subsection{Active Galaxy Sample}

Our first criterion for selecting a sample, and the only astrophysical one for the active sample, is to define a set of local AGN in a well-characterised way so that the selection effects are (as far as possible) understood.
The largest optical or infrared surveys that select AGN are often incomplete for nearby galaxies, and also tend to use AGN tracers that are anisotropic or impose a bias against star formation (by using large apertures to measure features that can be produced by both AGN and star formation).
For example, [O\,III]$\lambda5007$\AA, a typical optical line used in the selection of AGN, is not isotropic:
in comparison to [O\,IV]$\lambda25.9\mu$m, which correlates well with 10--200\,keV emission, it is underluminous in some Sy\,2s -- especially those that are Compton thick -- by up to an order of magnitude compared to Sy\,1s \citep{mel08,dia09,wea10}.
The same effect is seen in the 2--10\,keV emission when compared to the [O\,IV] line.
However, high ionisation lines may not be an ideal way to select AGN since they are not always observed:
\cite{kos13} detected Ne\,V only in 2/3 of luminous infrared galaxies for which an AGN was confirmed by the {\it Swift-BAT} survey.

In contrast to emission line and infrared tracers, the very hard 14--195\,keV band of the {\em Swift-BAT} survey measures direct emission from the AGN rather than scattered or reprocessed emission, and is much less sensitive to obscuration in the line-of-sight
than softer X-ray or optical wavelengths (selecting only against highly Compton thick AGN).
Indeed it is widely accepted as the least biased survey for AGN with respect to host galaxy properties, and as such it has been well studied.
There is a vast amount of ancillary data on the larger scale host galaxy properties \citep{kos10,kos11,win09,win10,mel14,mus14}, as well as analysis of the bolometric
corrections which enable one to make an estimate of the AGN luminosities \citep{vas10,win12,mel14}.
Because it is an all-sky survey with roughly uniform sensitivity, we can select a complete, volume limited sample of AGN.
Furthermore, the continuous nature of the survey means that more recent versions of the catalogue (58~and 70\,month, \citealt{bau13}) average over variability during the last 5--6\,years.

In addition to the single astrophysical criterion described above, we also impose two observational criteria to ensure homogeneous high resolution and observability. 
Our sample therefore consists of all 20 AGN in the 58\,month {\em Swift-BAT} catalogue that meet the following 3 criteria: 
(i) 14--195\,keV luminosities $\log{L_{14-195}} > 42.5$ (using redshift distance), 
(ii) redshift $z < 0.01$ (corresponding to a distance of $\lesssim40$\,Mpc), and 
(iii) observable from the VLT ($\delta < 15\deg$) so that they tend to be in the southern sky.
Note that the first two criteria are adjusted to intersect the flux limit of the 58-month catalogue for 90\% of the sky, that is $1.46\times10^{-11}$\,erg\,s$^{-1}$\,cm$^{-2}$.
The selection, and its completeness, is graphically represented in Fig.~\ref{fig:batselect}.

\begin{figure}
\epsscale{1.10}
\plotone{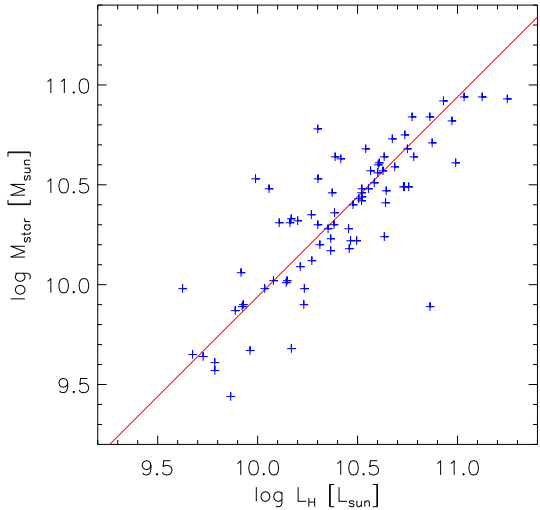}
\caption{\label{fig:mass_scale}
Comparison of the stellar mass $M_{star}$, derived from multi-band photometry, to the integrated H-band luminosity $L_H$ for a sample of AGN and other galaxies similar to that presented here \citep{kos11}. 
There is a clear relation with a scatter of $\sim0.2$\,dex, showing that $L_H$ can be used as a proxy for $M_{star}$.}
\end{figure}

\begin{deluxetable}{lcclcc}
\tabletypesize{\scriptsize}
\tablecaption{Summary of control sample properties\label{tab:control}} 
\tablewidth{0pt}
\tablehead{
\colhead{Name} &
\colhead{Distance} &
\colhead{log L$_{H}$} &
\colhead{Hubble} &
\colhead{axis} &
\colhead{bar} \\
\colhead{} &
\colhead{(Mpc)} &
\colhead{{L$_\odot$}} &
\colhead{stage} &
\colhead{ratio} &
\colhead{} \\
\colhead{(1)} &
\colhead{(2)} &
\colhead{(3)} &
\colhead{(4)} &
\colhead{(5)} &
\colhead{(6)} \\
}
\startdata

NGC 718      & 23 & \phn9.89 & \phm{\ \ -}1    & 0.87 & AB \\
NGC 1079     & 19 & \phn9.91 & \phm{\ \ -}0    & 0.60 & AB \\
NGC 1315     & 21 & \phn9.47 & \phm{\ \ }-1   & 0.89 &  B \\
NGC 1947     & 19 &    10.07 & \phm{\ \ }-3   & 0.87 &    \\
ESO 208-G021 & 17 &    10.88 & \phm{\ \ }-3   & 0.70 & AB \\
NGC 2775     & 21 &    10.45 & \phm{\ \ -}2    & 0.77 &    \\
NGC 3175     & 14 & \phn9.84 & \phm{\ \ -}1    & 0.26 & AB \\
NGC 3351     & 11 &    10.07 & \phm{\ \ -}3    & 0.93 &  B \\
ESO 093-G003 & 22 & \phn9.86 & \phm{\ \ -}0.3  & 0.60 & AB \\
NGC 3717     & 24 &    10.39 & \phm{\ \ -}3    & 0.18 &    \\
NGC 3749     & 42 &    10.40 & \phm{\ \ -}1    & 0.25 &    \\
NGC 4224     & 41 &    10.48 & \phm{\ \ -}1    & 0.35 &    \\
NGC 4254     & 15 &    10.22 & \phm{\ \ -}5    & 0.87 &    \\
NGC 4260     & 31 &    10.25 & \phm{\ \ -}1    & 0.31 &  B \\
NGC 5037     & 35 &    10.30 & \phm{\ \ -}1    & 0.32 &    \\
NGC 5845     & 25 &    10.46 & \phm{\ \ }-4.6 & 0.63 &    \\
NGC 5921     & 21 &    10.08 & \phm{\ \ -}4    & 0.82 &  B \\
IC 4653      & 26 & \phn9.48 & \phm{\ \ }-0.5 & 0.63 &  B \\
NGC 7727     & 26 &    10.41 & \phm{\ \ -}1    & 0.74 & AB \\

\enddata

\tablecomments{Columns: 
(1) common name;
(2) distances are, where possible, redshift independent (NED);
(3) integrated H-band luminosity (as a proxy for stellar mass).
(4) Hubble stage (NED);
(5) axis ratio (NED);
(6) B and AB denote presence of a bar or weak bar respectively (NED).}

\end{deluxetable}


\begin{figure*}
\epsscale{1.10}
\plotone{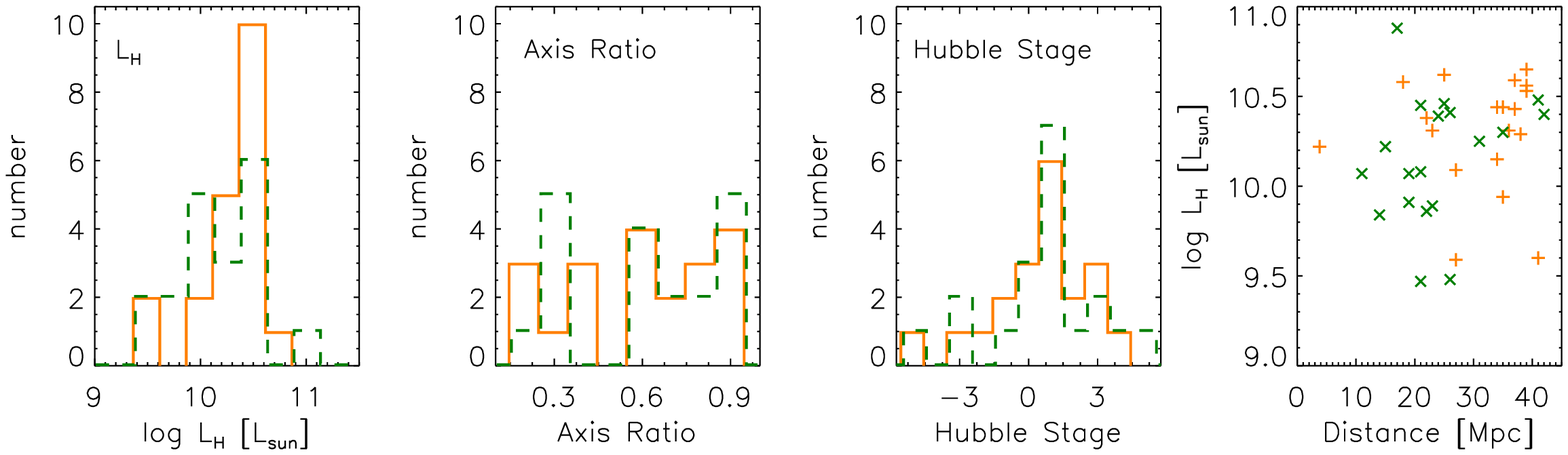}
\caption{\label{fig:prop_comp}
Comparison of several host galaxy properties between the hard X-ray selected AGN (orange; solid lines or plusses), and the matched inactive sample (dark green; dashed lines or crosses):
H-band luminosity as a proxy for stellar mass (far left), axis ratio (centre left), Hubble type (centre right), and H-band luminosity vs distance (far right).
There is no obvious relation between host galaxy luminosity and distance; and as expected by design, the distributions of the inactive sample properties match those of the AGN very well.
The biggest discrepancy is in distance, and this results from our preference to select closer, rather than more distant, inactive galaxies when possible.}
\end{figure*}

\subsection{Matched Inactive Sample}

The second criterion for our selection is that there should be a properly matched sample of inactive galaxies, so that it is possible to discern which features (stellar age, inflow rates, etc.) are related specifically to the AGN activity.
We describe the selection of the inactive sample below.
Although it has no part in the analysis presented in this paper in Sec.~\ref{sec:obsc}--\ref{sec:torus},  it does play a key role in many of the analyses which we will present in future papers.

The characteristics on which the inactive sample matching is based are: host galaxy morphology (Hubble type), inclination (axis ratio) and H-band luminosity (as a proxy for stellar mass).
For the purposes of target selection, we have derived the inclination from the axis ratio without any compensation for finite thickness.
This is not expected to yield a significant bias when comparing two galaxies with similar morphological type;
and the inclinations of the host galaxies will be assessed more carefully in future analyses.
The relation between H-band luminosity and stellar mass has been calibrated from a sample of similar galaxies for which the stellar mass has been derived using multi-band photometry \citep{kos11}.
Fig.~\ref{fig:mass_scale} shows that the scatter in this relation is 0.2\,dex.

We have quantified the matching criteria above using a $\chi^2$ metric with tolerances of $\pm1$ for Hubble type, $\pm15\deg$ for inclination, and $\pm0.3$\,dex for H-band luminosity.
In addition we require that the presence/absence of a bar is matched if possible. 
Finally, also where possible, we have selected inactive galaxies with redshifts less than or equal to those of their active pairs. 
All inactive galaxies were selected from the RC3 catalogue \citep{rc3}, rejecting any listed in the \cite{ver10} catalogue of known AGN. 
While some may still host very low luminosity AGN, the decisive factor is that (based on the lack of any obvious signature of BH acretion in optical, radio, and X-ray data) it is orders of magnitude weaker than in the X-ray selected sample.

The inactive galaxy selection is based on matching the characteristics of individual galaxies in the active sample.
This is used as a robust way to ensure that the samples, as a whole, match; it is not specifically our intention to compare pairs of active and inactive galaxies.
Part of the reason is that some inactive galaxies are a good match for several active galaxies, which means there is no unique pair matching.
The final matched inactive sample contains 19 galaxies, which are listed in Table~\ref{tab:control} together with the host properties used for the selection process.

A graphical comparison of the main host galaxy properties is given in Fig.~\ref{fig:prop_comp}.
By design the distributions of these properties should be similar, as is apparent from the figure.
This is confirmed for $L_H$, axis ratio, and Hubble stage, by Kolmogorov-Smirnov (KS) tests which indicate that the differences are not significant. 
Formally, the probabilities that the differences between the active and inactive samples could arise by chance exceed 20\%.
Both the active and inactive samples show the same distribution in H-band luminosity: the active sample has a mean of $\log{L_H [L_\odot]} = 10.3$ with a $1\sigma$ spread of 0.3 while the inactive sample has a mean of 10.2 and a spread of 0.4.
Both samples cover the full range of axis ratio (or equivalently inclination), with about half the sample each side of 0.7, equivalent to 45\deg\ for a disk.
And both samples cover a wide range of morphological types, with a clear preference for Hubble stage at $\sim1$ corresponding to S0 and Sa `early disk' types.
This distribution is consistent with the analysis presented by \cite{kos11}, who found for a larger sample of {\em Swift-BAT} AGN that only $\sim10$\% were in ellipticals, while $\sim30$\% had intermediate (S0) and $\sim40$\%  early type spiral (Sa--Sb) host galaxies.
\cite{kos11} noted that these fractions differ from the distribution of normal galaxies -- 16\% ellipticals, 26\% spirals.
But the distribution of host types of the {\em Swift-BAT} AGN are consistent with this once one takes into account the detection rate of Seyferts (rather than LINERs) reported by \cite{ho08}, which peaks for S0--Sb types.
\cite{dav14} discussed the role of host type and environment in the context of the origin of the gas fuelling the AGN, and argued that this has an impact on what gas inflow mechanisms one may see in the circumnuclear region.
This issue of morphology, including the presence or absence of a bar, will be revisited for our complete sample in a future work.
The main point here is that the difference in distributions underlines the need to include morphology as a criterion for matching the inactive sample as we have done.

The largest difference between the samples occurs for the distance distribution which a KS test indicates may be marginally significant at a level of 2.5$\sigma$.
This is a direct result of our requirement to select, whenever possible, inactive galaxies that are closer, rather than more distant, to their active pair.
We note that the active and inactive samples have mean distances of 31\,Mpc and 24\,Mpc respectively, and in both cases the $1\sigma$ spread is about 9\,Mpc.

\begin{figure*}
\epsscale{1.10}
\plotone{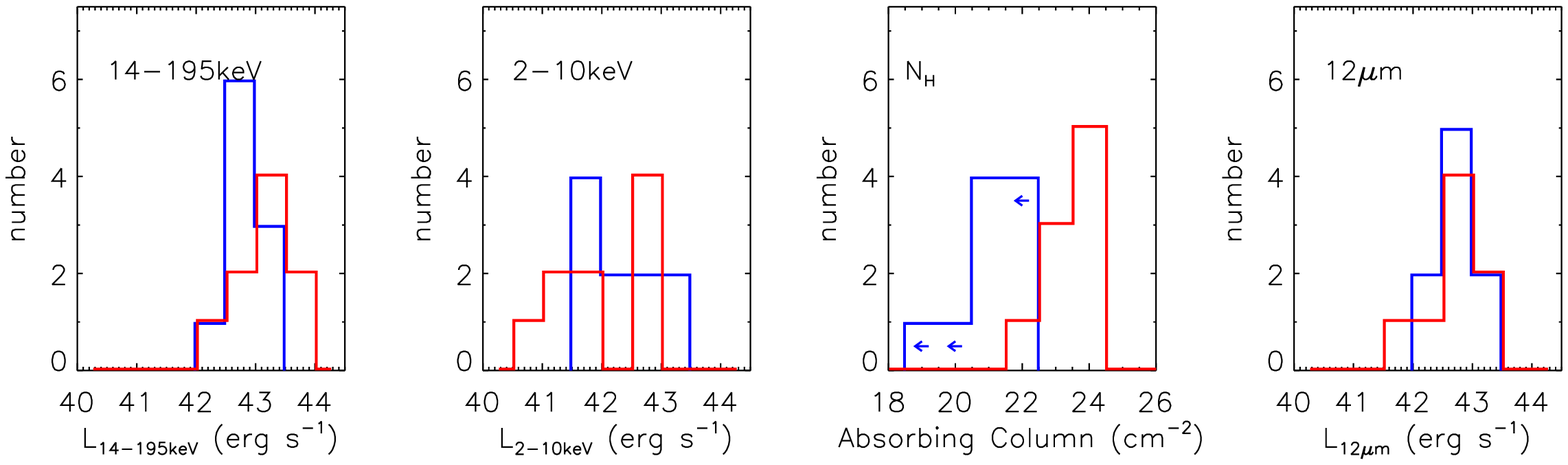}
\caption{\label{fig:prop_agn}
Comparison of nuclear properties of Sy\,1s (blue) and Sy\,2s (red).
There are no significant differences between Sy\,1s and Sy\,2s for any of the luminosity distributions shown. This holds even for the absorption corrected 14--195\,keV luminosity (not shown), despite the tantalising hint of a difference. In contrast, the difference in $\log{N_H}$ is 3.7$\sigma$.
Far left: the sample was selected according to 14--195\,keV luminosity, and shows no significant bias towards either Sy\,1s or Sy\,2s.
Centre left: Given the small numbers of objects, there is also no significant difference in the observed 2--10\,keV luminosities.
Center right: in contrast, and as expected, Sy\,2s do exhibit higher absorbing columns, with a median of $>10^{23}$\,cm$^{-2}$ in comparison to $\sim10^{21}$\,cm$^{-2}$ for the Sy\,1s (note we have marked AGN with no measurable neutral absorption as limits).
Far right: the nuclear mid-infrared luminosity shows a remarkably narrow distribution, similar to that of the 14--195\,keV luminosity, with no clear difference between Sy\,1s and Sy\,2s.}
\end{figure*}

\begin{figure*}
\epsscale{0.9}
\plotone{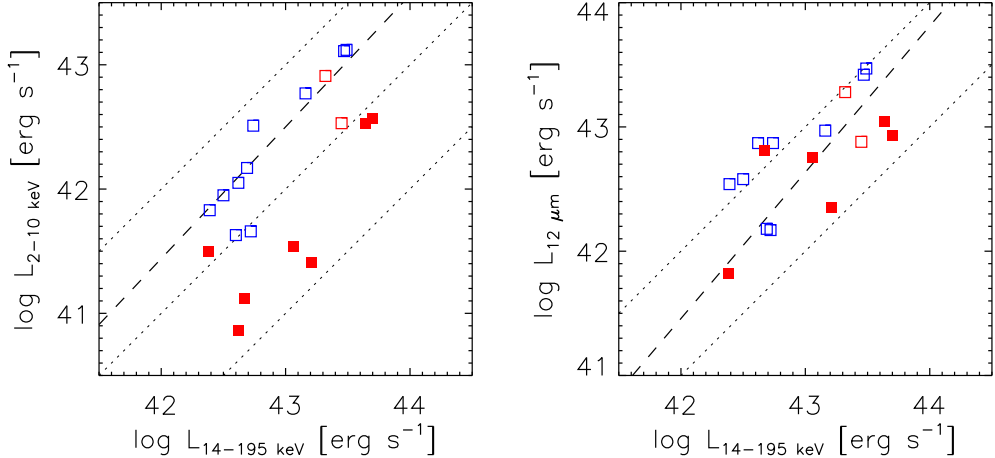}
\caption{\label{fig:xcorr}
Comparison of nuclear X-ray and mid-infrared luminosities of Sy\,1s and Sy\,2s (denoted by blue and red colour respectively).
Absorbing column is represented by open/filled symbols: filled squares have $N_H \geq 10^{23}$\,cm$^{-2}$ while open squares have $N_H < 10^{23}$\,cm$^{-2}$ (as discussed in Sec.~\ref{sec:agntype}, this threshold is where $N_H$ starts to have an impact on the 2-10\,keV band).
The two red open squares are both AGN classified as Sy\,2\,(1i). 
\cite{bur15} argue that, based on their physical properties, these should be considered more like Sy\,1 than Sy\,2.
Left: $L_{14-195}$ vs observed $L_{2-10}$. 
From top left to bottom right, the dotted lines trace ratios of $L_{14-195}/L_{2-10} = 1$, 10, 100 (indicative of increasing absorbing column, see \citealt{kos13}).
The relation of \cite{win09} for $L^{int}_{2-10}$ is indicated by the dashed line.
Sy\,1s and other AGN with low N$_H$ follow this relation, indicating that for these $L_{2-10} \sim L^{int}_{2-10}$.
Right: $L_{14-195}$ vs nuclear  $L_{12\mu m}$. 
From top left to bottom right, the dotted lines indicate ratios of $L_{14-195}/L_{12\mu m} = 1$, 10.
The dashed line denotes the relation of \cite{gan10}, replacing $L^{int}_{2-10}$ with $L_{14-195}$ according to \cite{win09}.
To first order, the data follow this relation without any dependency on $N_H$ or AGN type.
}
\end{figure*}

\section{AGN properties}
\label{sec:agntype}

In this section we compare various properties of the Sy\,1s and Sy\,2s in the active sample, as listed in Table~\ref{tab:agn}, focussing on the absorbing column and its impact.
We will, as our data become available, update the classifications of the AGN if necessary.
Similarly, and as applicable to the specific analyses we will perform, we will re-assess the host galaxy properties of the active and inactive samples: morphological classifications, whether they are barred, and their inclinations.

The sample was selected purely on $L_{14-195}$.
Adopting a simple conversion to estimate the bolometric AGN luminosity as \citep{win12}
\[
\log{L_{AGN}} = 1.12 \log{L_{14-195}} - 4.23
\] 
(i.e. bolometric corrections of 5--10 for the luminosity range here)
and without applying any absorption correction (see below), our complete volume limited sample has a median luminosity of $\log{L_{AGN}} = 43.5$ with a $1\sigma$ distribution of 0.5\,dex.

Table~\ref{tab:agn} shows that our selection has yielded similar numbers of Sy\,1s and Sy\,2s, and the first panel of Fig.~\ref{fig:prop_agn} indicates that any difference in their $L_{14-195}$ distributions is not significant.
This suggests that, for the number of objects in the sample, any preference to select either Sy\,1s or Sy\,2s is not significant.
For both $L_{14-195}$ and $L_{12\mu m}$, a KS test indicates that the probability of the difference between the Sy\,1 and Sy\,2 distributions arising by chance exceeds 45\%.
The difference for the observed $L_{2-10}$ is also not enough to be significant.
It does, however, have a measurable impact on the ratio of the observed 14--195\,keV and 2--10\,keV luminosities, which is 4 for the Sy\,1s and 17 for the Sy\,2s.
The same characteristic is clearly seen in the left panel of Fig.~\ref{fig:xcorr} where $L_{14-195}$ is plotted versus observed $L_{2-10}$.
Here Sy\,1s/2s are represented by blue/red colour respectively.
The Sy\,1s follow the relation between $L_{14-195}$ and $L^{int}_{2-10}$ derived by \cite{win09}
\[
\log{L^{int}_{2-10}} \, = \, 1.06 \, \log{L_{14-195}} \, - \, 3.08,
\] 
(in the luminosity range here, this relation is approximately equivalent to $\log{L^{int}_{2-10}} \sim \log{L_{14-195}} - 0.5$) while many of the Sy\,2s do not.
The reason is due to their different absorbing columns $N_H$, which is denoted by open/filled symbols in Fig.~\ref{fig:xcorr}.
And indeed, the observed $L_{14-195}/L_{2-10}$ ratio can in principle be used to give an estimate of $N_H$ \citep{kos13}.

The difference in $N_H$ is clearly seen in the centre right panel of Fig.~\ref{fig:prop_agn}, and a KS test gives it a significance of 3.7$\sigma$.
Values for $N_H$ are taken from \cite{ric15}, who derived absorbing columns for many of the {\it Swift-BAT} AGN in a consistent way, fitting the 0.3--150\,keV spectral energy distribution.
The basic model includes an absorbed cutoff power-law continuum plus a reflection component (for details see \citealt{ric15}).
For sources with $N_H \lesssim 10^{22}$\,cm$^{-2}$, they included various additional components if statistically needed: 
a blackbody (for the soft excess), 
partially covering ionized absorption, 
a cross-calibration constant (for possible variability between the non-simultaneous soft X-ray and {\em Swift-BAT} observations),
and an iron line (or other emission lines in that region). 
Similarly, for the more obscured sources, some additional components were added if required: 
a scattered component, a collisional plasma, emission lines, and a cross-calibration constant. 
Note that in the column densities given in Table~\ref{tab:agn} we do not include warm absorbers, which arise in ionised out-flowing gas and are observed in about 50\% of Sy\,1s \citep{kom99,blu05,win12}.
The difference between the neutral absorption in Sy\,1s and~2s is significant.
The mean for the Sy\,1s is $\log{N_H [cm^{-2}]} \sim 21.1$ while for Sy\,2s it is $\log{N_H [cm^{-2}]} = 23.5$.

Models of the propagation of X-rays through gas in various geometries around a central source \citep{mat99,bri11,yaq12,ric15} indicate that $N_H$ can account for the different luminosity ratio between the Sy\,1s and Sy\,2s.
These models indicate that the flux correction for the 14--195\,keV band is negligible up to 10$^{23}$\,cm$^{-2}$, and reaches a factor 2 around 2--$3\times10^{24}$\,cm$^{-2}$, and a factor 10 at 10$^{25}$\,cm$^{-2}$ (whether higher corrections for larger columns are required depends on the reflection component).
As such, even the absorption corrected 14--195\,keV luminosities do not, based on a KS test, yield a significant difference between Sy\,1s and Sy\,2s.
On the other hand, large flux corrections, greater than a factor 2, to $L_{2-10}$ are already required for absorbing columns of $\sim10^{23}$\,cm$^{-2}$, increasing to a factor 10 at $\sim5\times10^{23}$\,cm$^{-2}$.
If not accounted for, this can lead to a bias when selecting via the 2--10\,keV band, since the impact on the observed flux means that some Sy\,2s may be excluded from a sample even before such a correction can be made.

This effect is much reduced when selecting in the 14-–195\,keV band.
Its impact is apparent in the distribution of N$_H$ for the Sy\,2s: as expected from \cite{ris99}, most of the sources have N$_H \gtrsim 10^{23}$\,cm$^{-2}$.
And (allowing for variations in N$_H$) 3--5 of the sources have N$_H > 10^{24}$\,cm$^{-2}$, consistent with the expectation that 10--30\% of AGN are Compton thick \citep{tre09,ale13}.
But in contrast to \cite{ris99} there are none with N$_H \gtrsim 10^{25}$\,cm$^{-2}$, which is simply due to the absorption in the band at such high columns.
However, we expect very few local sources to fall in this regime.
NGC\,1068 is the best known example of one that does, and hence is excluded from the sample: although its $L_{14-195}/L_{2-10} \sim 10$ is not immediately indicative of a high column, early estimates \citep{mat97} suggested it is highly Compton thick, and fits to recent hard X-ray data require several components with columns up to 10$^{25}$\,cm$^{-2}$ \citep{bau14}.

The final comparison is of the nuclear 12$\mu$m luminosities, taken from \cite{asm14}.
These data were observed at subarcsecond resolution, and so resolve out mid-infrared continuum due to star formation in the circumnuclear region.
\cite{gan10} and \cite{asm12} have shown that there is a very good correlation between $L^{int}_{2-10}$ and $L_{12\mu m}$, with no major difference between Sy~1s and Sy~2s.
Our comparison in Figs.~\ref{fig:prop_agn} and~\ref{fig:xcorr} confirms that this is also the case for our sample, that there is little dependence on $L_{12\mu m}$ or its relation to $L_{14-195}$ with either AGN type or $N_H$. 
Note, however, \cite{bur15} point out that the nuclear 12$\mu$m continuum does show some signs of a slight anisotropy, which can also be seen in the right panel of Fig.~\ref{fig:xcorr}.

\section{Optical Obscuration at Low Luminosity}
\label{sec:obsc}

The proportion of AGN with substantial optical obscuration and/or X-ray absorption, and their dependence on luminosity, offers insight into the properties of the obscuring structures around the AGN.
In this section we focus on low luminosities, in the range 
$42.5 < \log{L_{14-195}} < 44$.
After clarifying in Sec.~\ref{sec:def} the definitions we use for optically obscured and X-ray absorbed AGN, and summarising the various types of AGN that are found, we discuss X-ray unabsorbed Sy\,2s in Sec.~\ref{sec:type21s}, and put this in the context of the {\em Swift-BAT} sample in Sec.~\ref{sec:simBAT} to assess how rare such objects really are.

\subsection{Definitions and Classifications}
\label{sec:def}

\begin{figure}
\epsscale{0.9}
\plotone{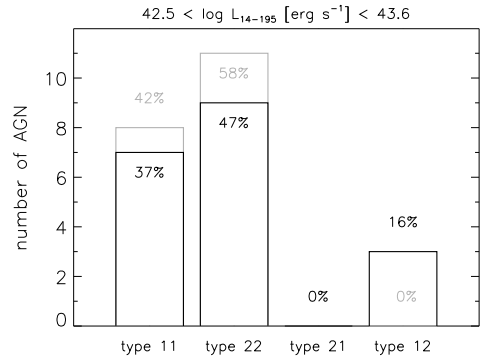}
\caption{\label{fig:battype}
Fractions of optically obscured and X-ray absorbed AGN in our complete volume limited {\it Swift-BAT} sample (see also Fig.~\ref{fig:syfrac}). 
The first/second digit of the type codes for optical/X-ray obscuration as described in Sec.~\ref{sec:def}.
The black lines denote the strict definitions.
The grey lines show the impact of relaxing the definitions slightly: the three type 12 AGN would be reclassified as one type 11 and two type 22.}
\end{figure}

\begin{figure}
\epsscale{1.10}
\plotone{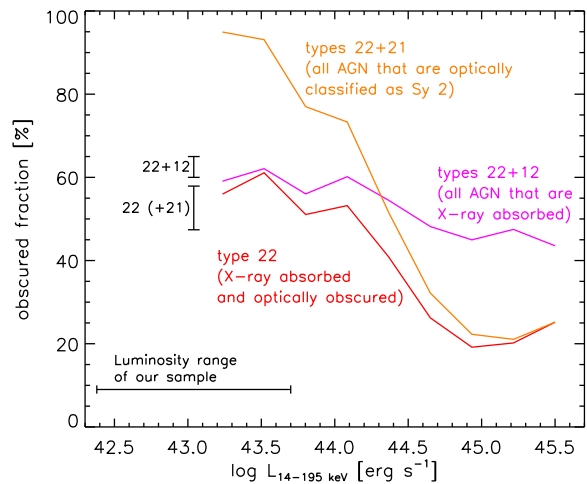}
\caption{\label{fig:syfrac}
Fractions of optically and X-ray obscured AGN as a function of luminosity.
The curves are adapted from Fig.~12 of \cite{mer14}. The details of the types, defined by those authors, are summarised in Section~\ref{sec:simBAT}: the first/second digit of the type codes for optical/X-ray obscuration.
The luminosity scale has been derived from $L^{int}_{2-10}$ as indicated in Section~\ref{sec:agntype}.
The black ranges refer to our sample, and indicate the impact of allowing some flexibility in the definition of optically obscured and X-ray absorbed.
At overlapping luminosities, the sample have similar fractions of AGN that are X-ray absorbed (types 22+12) and that are both X-ray absorbed and optically obscured (type 22); but the fractions of AGN classified as Sy\,2 (types 22+21) are very different.}
\end{figure}

Much of the discussion below is set in the context of \cite{mer14}.
As such, we follow these authors and adopt N$_H>10^{21.5}$\,cm$^{-2}$ as the criterion for defining an AGN to be X-ray absorbed.
For typical gas-to-dust ratios this is equivalent to an optical screen extinction of A$_V\sim2$\,mag, sufficient to mildy obscure the line of sight to the broad line region (BLR) at visible wavelengths\footnote[2]{We do not distinguish a category of obscuration in which the broad emission lines are obscured at optical wavelengths but can be detected at near-infrared wavelengths. These objects, which are sometimes classified as Sy\,1i, are considered to be Sy\,2s in this work. We note, however, that like Sy\,1.8--1.9s they are close to the boundary of optically obscured/unobscured and X-ray absorbed/unabsorbed. And there are good reasons to consider them instead as Sy\,1 as discussed by \cite{bur15}.}.
In our volume limited 14--195\,keV sample, we find 50--60\% of the AGN are optically obscured while $\sim60$\% are X-ray absorbed.
Most of the obscured AGN are both optically obscured and X-ray absorbed;
none are just optically obscured;
and only three are just X-ray absorbed.
These exceptions are `minor' since they are all close to the boundary between the regimes.
The X-ray absorbed AGN NGC\,1365 is optically classified as Sy\,1.8 and, although in many studies is taken to be a Sy\,2, here we adopt a strict definition and so classify it as Sy\,1.
Similarly MCG-05-23-016 is X-ray absorbed and has only weak broad H$\alpha$, hence its previous classification as a Sy\,2 which we have revised to Sy\,1.9.
The Sy\,1.8 NGC\,2992 has a column only marginally above our X-ray absorption threshold and so is on the borderline in both classifications.
It would be classified as X-ray unabsorbed according to other definitions that set the threshold at $10^{22}$\,cm$^{-2}$ \citep{pan02}.
As such, there could be two more optically obscured and one less X-ray absorbed AGN than found using the strict definitions we apply in Fig.~\ref{fig:battype}.
This is reflected by the ranges shown in Fig.~\ref{fig:syfrac}.

The joint optical/X-ray classifications are summarised in Fig.~\ref{fig:battype} where we adopt the terminology of \cite{mer14} in which the first/second digit of the classification denotes the optical/X-ray type.
This leads to the following types:
\begin{description}
\item[type 11] optically unobscured and X-ray unabsorbed;
\item[type 22] optically classified as Sy\,2 and X-ray absorbed;
\item[type 21] optically classified as Sy\,2 but X-ray unabsorbed; 
\item[type 12] optically unobscured but X-ray absorbed;
\end{description}
Thus, with reference to the curves in Fig~\ref{fig:syfrac} adapted from \cite{mer14} to the 14--195\,keV luminosity scale, we also have:
\begin{description}
\item[type 22+12] all X-ray absorbed AGN (regardless of their optical classification); 
\item[type 22+21] all AGN optically classified as Sy\,2 (independent of whether they are X-ray absorbed).
\end{description}

Comparing the AGN in our sample to the left edge of Fig~\ref{fig:syfrac} immediately highlights a discrepancy.
Above, we noted that 50--60\% of our complete sample are optically obscured and classed as Sy\,2s, a fraction similar to that estimated by \cite{law10}.
However, \cite{mer14} class $\sim90$\% of their AGN in an overlapping luminosity range $\log{L_{14-195}} \sim 43$--43.5 as Sy\,2s.
The difference is due to the large fraction of AGN classified by these authors as type~21, that is X-ray unabsorbed Sy\,2 galaxies \citep{pap01,pan02,bri08,bia12}.

\subsection{How common are X-ray unabsorbed Seyfert 2s?}
\label{sec:type21s}

X-ray unabsorbed Sy\,2s are believed to be galaxies with a direct view to the AGN but in which there is no BLR (such objects are also known as pure or true Sy\,2s).
This view is supported by, for example, the 6 objects identified by \cite{haw04} as having optical spectra typical of Sy\,2s (H$\beta$ FWHM $<$ 1000\,km\,s$^{-1}$ and [O\,III]/H$\beta$ $>$ 3) but large amplitude variations typical of Sy\,1s.
Subsequent X-ray observations of three by \cite{gli07} confirmed that these do not have significant absorption in the 0.3--8\,keV band.
These authors also showed that these AGN are relatively luminous (intrinsic $\log{L_{0.5-8}}>43.2$).
Thus, despite their rather low Eddington ratios $L_{bol}/L_{Edd}<0.01$ \citep{gli07}, the absence of a BLR cannot be explained by low luminosities and/or accretion rates either being unable to sustain a disk wind \citep{nic00,eli09} or meaning that high dispersion prevents BLR clouds from surviving \citep{lao03}.
At present, the reason that they may not have a BLR is still open.

These objects may be related to the Sy\,2s for which spectropolarimetry shows no evidence of a hidden BLR.
There have been many studies of polarised emission from hidden BLRs. 
The largest indicate that Sy\,2s in which a hidden BLR has not been detected have less luminous AGN than other Sy\,2s but are not more optically obscured \citep{tra01,gu02,tra03} -- qualitatively, but not necessarily quantitatively, consistent with the idea that BLR may not form in low luminosity AGN.
It also suggests that a BLR is completely absent in at least 50\% of Sy\,2s.
However, this fraction is difficult to put in perspective because of complications due to the impact of detection limits as well as the expected scattering efficiency.
The latter issue was addressed by \cite{hei97} who showed that the ability to detect polarised emission was related to the far-infrared colours in a way that suggested the scattering particles were in the throat of the torus, and so even they would be hidden at high inclinations.
Following on from work of \cite{ram11} which shows that the intrinsic properties (rather than just the inclination) of the tori in Sy\,1s and Sy\,2s are different, \cite{ich15} have fitted torus models to Sy\,2s with and without polarised emission.
Their results again indicate that whether a hidden BLR is seen depends on the location of the scattering material in the throat of the torus; but that rather than inclination, it is the opening angle and covering factor of the torus that determines this difference.
As such, the fraction of Sy\,2s in which polarised scattered BLR emission is not observed could be very different from the fraction without a hidden BLR.
Since, based on these models, the polarised BLR emission may in some cases be only relatively lightly obscured, deep and/or near-infrared spectropolarimetry may shed further light on this issue.
NGC\,7172 and NGC\,7582 (see Table~\ref{tab:agn}) are cases in point here: \cite{bia07} and \cite{mar12} both report that no polarised broad H$\alpha$ has been detected but \cite{sma12} have seen broad Pa$\alpha$ and Br$\gamma$ directly in NGC\,7172 and \cite{reu03} broad Br$\gamma$ in NGC\,7582.

Despite the uncertainties, the evidence that at least some relatively luminous AGN do not have a BLR is convincing. 
The key question for this paper is how common such objects are, since this can have a significant impact on the total fraction of AGN optically classified as Sy\,2.

In their analysis, \cite{ris99} found that about 4\% of Sy\,2 galaxies had $N_H < 10^{22}$\,cm$^{-2}$.
However, \cite{pan02} presented a sample of 17 objects which had been classified as Sy\,2 in the literature, and which had similarly low column densities.
They suggested that X-ray unabsorbed Sy\,2s may be much more common, being 10--30\% of all Sy\,2s.
More recently, other authors have confirmed that such objects do exist, but without a consensus on how common they are, because many candidates are ruled out on closer examination.
For example, \cite{bri08} found that of 8 candidates, 4 were underluminous in the 2--10\,keV bands suggesting that they were Compton thick -- with the unabsorbed softer X-rays originating in a scattered continuum or from host galaxy contamination.
Similarly, in an examination of 8 candidates -- including some from the \cite{pan02} sample -- \cite{bia12} ruled out 4 for a variety of reasons, and confirmed only 3 as unabsorbed Sy\,2s.
Using a variety of metrics for a critical assessment of 24 objects that were previously claimed to be unabsorbed Sy\,2s (again including many from \citealt{pan02}), \cite{shi10} could confirm only two as genuine, with broad emission lines 2--3 orders of magnitude fainter than typical Sy\,1s -- although several more candidates may have anomalously weak broad lines, this could not be confirmed with existing data.
These authors concluded that 1\% or less of Sy\,2s are X-ray unabsorbed.

In contrast, a much higher fraction of such sources -- 30\% of AGN at $\log{L_{14-195}} \sim 43$ -- appears in the \cite{mer14} sample.
Whether this indicates that there is a large population of X-ray unabsorbed Sy\,2s or is due to a classification bias was discussed in depth by these authors.
With reference to stacked optical spectra and spectral energy distributions, they suggested that many of the lower luminosity AGN may have been incorrectly photometrically classified as Sy\,2 due to the low contrast of the AGN against the relatively bright host galaxy.
On the other hand, spectroscopic data are sensitive to features associated with Sy\,1s, such as broad lines, even when they are weak.
Such signatures would not be evident in broad-band integrated photometry.
This difference is reflected in their data: while 77\% of the AGN with only integrated broad-band photometric data are classified as Sy\,2, only 56\% of AGN with spectroscopic data are classified as Sy\,2s.
The lower total fraction of Sy\,2s based on spectroscopic classifications implies in turn a much lower fraction of X-ray unabsorbed Sy\,2s.

Fig.~\ref{fig:battype} shows that in our volume limited sample of {\em Swift-BAT} AGN, there are no sources classified as X-ray unabsorbed Sy\,2, implying these sources are rare.
In this context, a simulation of the AGN population can provide valuable insight, since we know the Sy\,2 fractions in both flux limited and volume limited samples.
We describe this simulation in Sec.~\ref{sec:simBAT}, aiming to answer the question of which curve from \cite{mer14} in Fig.~\ref{fig:syfrac} properly traces the fraction of Sy\,2s: is it types 22+21 or just type 22?

\begin{figure*}
\epsscale{1.10}
\plotone{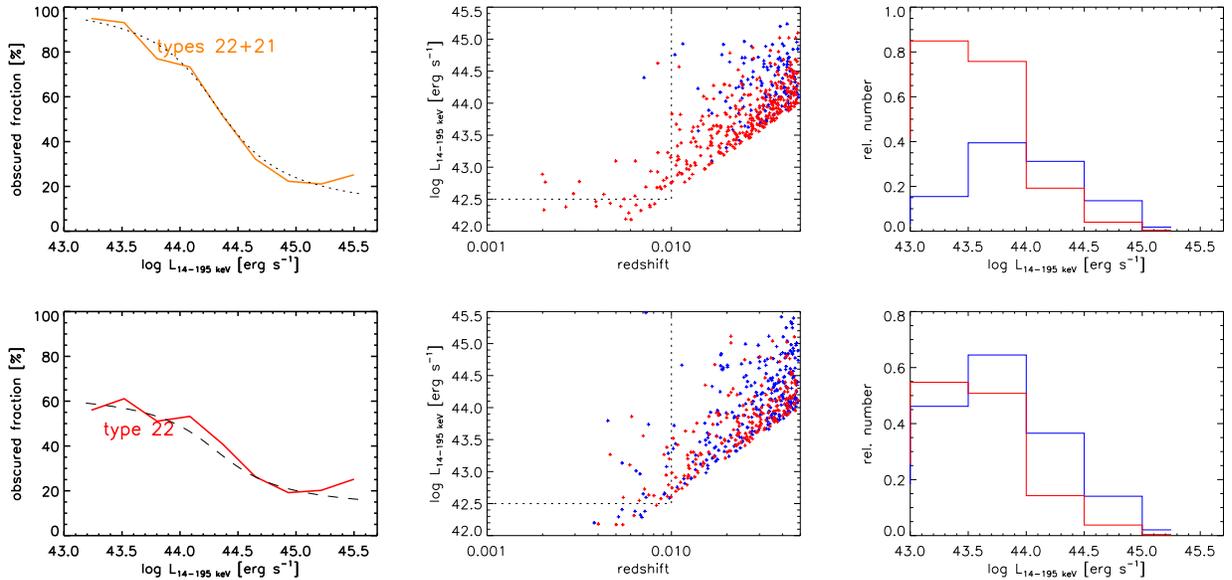}
\caption{\label{fig:fracsim}
Simulations of the {\it Swift-BAT} AGN sample using the 2 different optical obscuration fractions, as described in Section~\ref{sec:simBAT}. Top panels: using the fraction of AGN classified as Sy\,2 (type `22+21'); bottom panels: using the true fraction of optically obscured AGN (type `22'). 
Left panels are from Fig.~\ref{fig:syfrac}, showing the fraction of AGN classified by \cite{mer14} as Sy\,2 (types 22+21, top) and the Sy\,2s that are also X-ray absorbed (type 22, bottom).
The dotted line (top) indicates a functional approximation to the `types 22+21' curve from \cite{mer14}; the dashed line (bottom) is our modification to match the `type 22' curve.
Centre panels show each AGN in the luminosity-redshift plane (for clarity only a random subset is drawn). The flux limit is clearly apparent, and the thresholds for our complete sample are marked. These panels can be compared to Fig.~\ref{fig:batselect}. Right panels: histograms showing the distributions of AGN luminosity in a flux limited sample, that can be compared with Fig.~\ref{fig:win09fig7} which is redrawn from Fig.~7 in \cite{win09}. In all panels, Sy\,1s are drawn in blue and Sy\,2s in red.}
\end{figure*}

\begin{figure}
\epsscale{1.10}
\plotone{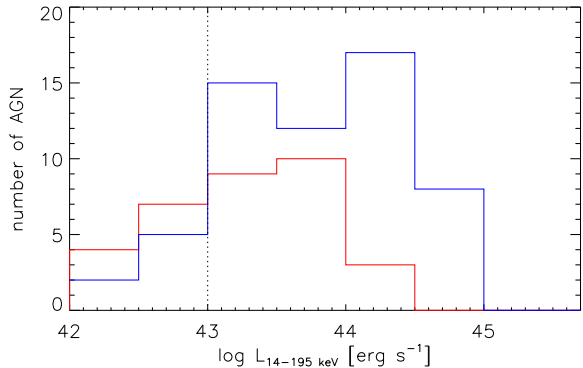}
\caption{\label{fig:win09fig7}
Number of Sy\,1s (blue) and Sy\,2s (red) in the flux limited sample of \cite{win09}. This figure is redrawn from their Fig.~7 using data from their Tables~1 and~4, but plotted as a function of 14--195\,keV luminosity derived from $L^{int}_{2-10}$ as indicated in Section~\ref{sec:agntype}. The luminosity range to the right of the dotted line is that discussed in this paper.}
\end{figure}

\subsection{A Simulated Swift-BAT sample}
\label{sec:simBAT}

Our aim here is to simulate parent populations of AGN with different prescriptions for the intrinsic fraction of Sy\,2s.
After applying observational limits, we can then use constraints from the flux-limited sample of \cite{win09} and our own volume-limited sample to discriminate between the different prescriptions, giving us a handle on the true fraction of Sy\,2s.

We begin by constructing a population of AGN using a 2--10\,keV luminosity function from \cite{air10} suitable for low redshift:
\[
\phi(L^{int}_{2-10}) \ \propto \ [(L^{int}_{2-10}/L_0)^{0.70} + (L^{int}_{2-10}/L_0)^{3.14}]^{-1}
\]
where $\log{L_0} = 44.96$.
We have then randomly assigned each AGN to be a Sy\,1 or Sy\,2 according to the obscured fraction as a function of luminosity. 
We use 2 functions, which are discussed below.
We allow the AGN to be observed by calculating $L_{14-195}$ from $L^{int}_{2-10}$, assigning them random locations in a local volume, deriving their fluxes, and assessing whether they would be detected in the {\it Swift-BAT} survey.
In doing this, we have not included the effects of X-ray absorption in the 14--195\,keV band since, as we have seen, it has little impact at such high energies for the absorbing columns expected.
We use a limit appropriate for the 9-month survey in order to select a flux limited sample matching that used by \cite{win09};
and a limit appropriate for the 58-month survey to match our volume limited selection outlined in this paper.

The left panels in Fig.~\ref{fig:fracsim} show the 2 obscuration functions used in our simulations.
The dotted line in the upper left panel is taken directly from \cite{mer14} and was designed to follow the curve for types 22+21; 
the dashed line in the lower left panel is our modification of this function to match the type 22 data, those AGN that are both optically obscured and X-ray absorbed.
The results of our simulations using these functions are also given in Fig.~\ref{fig:fracsim}.
The centre panels show the Sy\,1 and Sy\,2 populations generated -- which can be compared to the real data in Fig.~\ref{fig:batselect}.
The right panels show their resulting distributions as a function of luminosity, which can be compared to Fig.~\ref{fig:win09fig7}.
In these panels, the sharp decrease in the number of Sy\,2s around $\log{L_{14-195}} \sim 44$ simply reflects the rapid change in the fraction of obscured AGN at this luminosity.
The flux limit means that low luminosity AGN are not detected at larger distance, and hence biasses the apparent fraction of obscured AGN at $\log{L_{14-195}} < 43.5$.
In contrast, because our sample is complete to $\log{L_{14-195}} = 42.5$, it provides an unbiassed indication of the obscured fraction at these luminosities.

It is immediately clear that the lower panels better match the observed distribution of Sy\,1 and Sy\,2 sources reported in \cite{win09}.
Similarly, the lower panels correspond to finding a fraction 54\% of Sy\,2s (in contrast to 87\% for the upper panels) in a complete volume limited sample equivalent to that described in Sec.~\ref{sec:sample}, again providing a good match to the fraction actually measured in the {\it Swift-BAT} sample.

Our conclusion is that, over the luminosity range $43 < \log{L_{14-195}} < 44.5$ where the `type 22' and `types 22+21' curves differ in Fig.~\ref{fig:syfrac}, the {\em true} fraction of Sy\,2s is given by the `type 22' curve. 
This confirms that X-ray unabsorbed Sy\,2s are rare, at most a few percent of the Sy\,2 population; 
and that the high number of type~21 objects in the \cite{mer14} sample is due to a bias in photometric classification.
It also implies that, in low luminosity systems, optical obscuration and X-ray absorption are usually found together -- and hence most likely originate in the same shared obscuring structure.

\section{The Torus and Broad Line Region}
\label{sec:torus}

\begin{figure*}
\epsscale{1.0}
\plotone{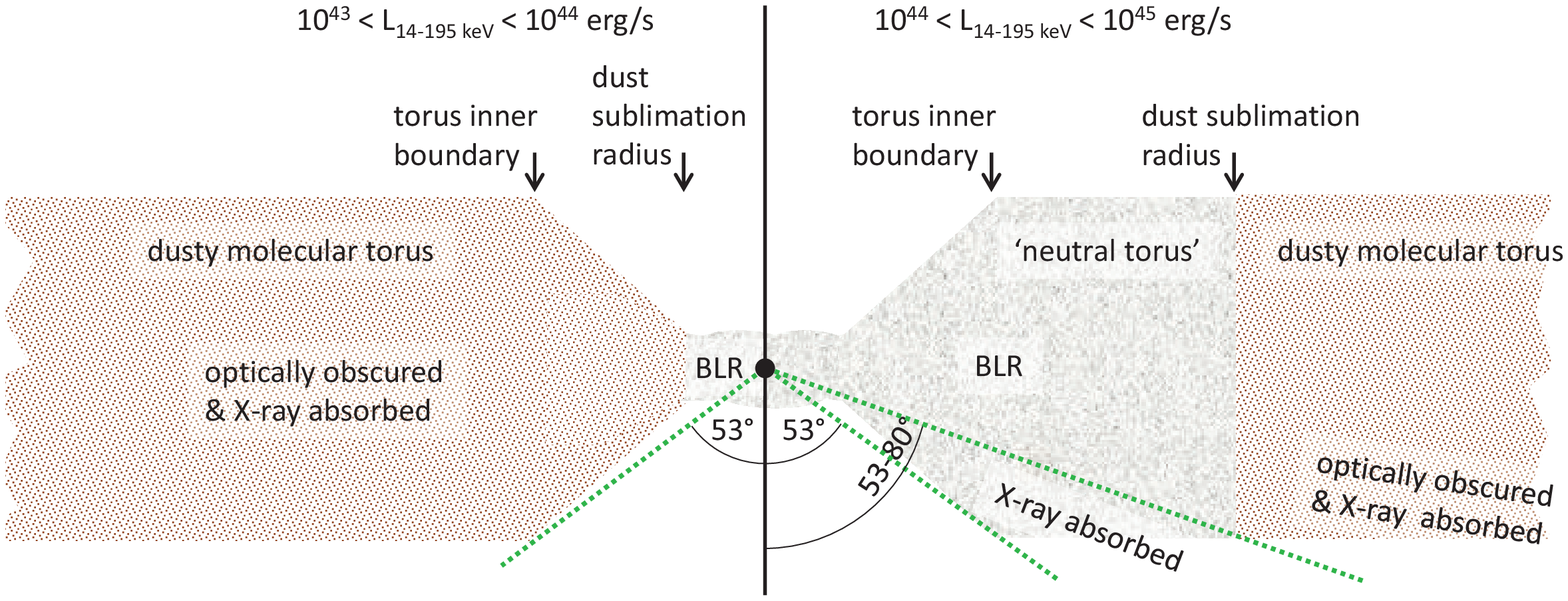}
\caption{\label{fig:cartoon}
Cartoon illustrating how the inner edge of the torus and the dust sublimation radius $R_{dust}$ are related, depending on luminosity (we do not distinguish smooth or clumpy torus concepts here, and the cartoon could also be envisaged as a less precisely demarcated cloud distribution).
At the luminosity range on the left side, the fraction of optically obscured and X-ray absorbed AGN is the same.
Because $R_{dust}$ is smaller than the radius of the inner edge of the geometrically thick dusty torus, it has no impact on the obscured fraction (which is set by the height of the thick structure).
The right side depicts the situation at higher luminosities.
Here, one finds essentially the same geometrically thick gas structure, but now $R_{dust}$ is larger than the radius of its inner wall.
Thus the broad line region extends outward into the neutral torus which, further out still, becomes the dusty torus.
At these luminosities, the fraction of X-ray absorbed AGN is the same, but the optically obscured fraction decreases with increasing luminosity.
}
\end{figure*}

Fig.~\ref{fig:syfrac} shows that the behaviour of the Sy\,2 fraction and X-ray absorbed fraction of AGN are different above and below a luminosity of $\log{L_{14-195}} \sim 44$.
In this section we explore what this may indicate about the torus.
Often, the torus is simply considered to be a dusty entity, since it must obscure the BLR in some AGN.
A more careful definition would be that the torus is a geometrically thick structure (since it confines the ionisation cone and any outflow) that causes optical obscuration and some (perhaps much) of the cold X-ray absorption.
We show that, in this case, its inner boundary is decoupled from the dust sublimation radius.
Thus, while near-infrared reverberation mapping \citep{sug06} and interferometric data \citep{kis09,kis11,wei12} measure the dust sublimation radius, they do not measure the inner boundary of the torus.

\subsection{At low luminosity, $\log{L_{14-195}} < 44$}
\label{sec:lowlum}

At luminosities below $\log{L_{14-195}} \sim 44$, the optically obscured AGN fraction of $\sim60$\% is similar to the X-ray absorbed fraction, and is independent of luminosity.
Since the dust sublimation radius $R_{dust}$ increases with luminosity, this independence must reflect the geometrical properties of the torus which we can constrain as follows.
A luminosity of $\log{L_{14-195}} = 44$ implies $R_{dust,44} \sim 0.2$\,pc for graphite grains of size 0.05\,$\mu$m \citep{net14}.
An obscured fraction of 60\% implies $H/R \sim 0.75$ -- a half opening angle of $\sim53\deg$ -- and hence $H_{dust,44} \sim 0.15$\,pc.
And that the obscured fraction is constant with luminosity implies that $H/R$ cannot increase at radii smaller than $R_{dust,44}$ (and the rapid drop in the optically obscured fraction at higher luminosities indicates that $H/R$ also decreases at larger radii out to a few times $R_{dust,44}$).
This is illustrated in the left side of Fig.~\ref{fig:cartoon}.
There are two implications.
First, it is the same geometrically thick structure that causes the optical obscuration and the X-ray absorption, and this structure is synonymous with the dusty torus.
The implied half-opening angle of $\sim53\deg$ is at the upper end of the 30--50\deg\ range derived from 3D models of the ionisation cone kinematics \citep{mue11,fis13}, a marginal consistency that could reflect uncertainties in the derivations of these numbers or may have a more physical origin.
Secondly, the inner wall of the dusty torus -- which defines the optically obscured and X-ray absorbed fractions -- has universal properties at $\log{L_{14-195}} < 44$: the average torus properties are an inner radius of $\sim0.2$\,pc and a scale height of $\sim0.15$\,pc, independent of luminosity.
Here we make two notes.
The first is that this is a constraint on the geometrically thick nature of the torus, and any dust that exists at smaller radii will not contribute to the obscured fraction because it is not driving the maximum value of $H/R$ \citep{net93,net14}.
The second is that at luminosities below that of our sample, one would expect the torus properties to change if there is a luminosity threshold below which the torus disappears \citep{eli06}.

\subsection{At high luminosity, $\log{L_{14-195}} > 44$}
\label{sec:highlum}

While none of the AGN in our sample are this luminous, the data of \cite{mer14}, as reproduced in Fig.~\ref{fig:syfrac}, show a dramatic difference in the Sy\,2 and X-ray absorbed fractions.
At these luminosities the Sy\,2 fraction is not expected to suffer a significant bias due to mis-classification simply because the AGN are brighter with respect to the host galaxy; and indeed the fraction of AGN classified as type 21 by \cite{mer14} does decrease dramatically.
As such, one can consider the Sy\,2 fraction as the optically obscured fraction.
This fraction drops rapidly to $\sim20$\% following the $L^{0.5}$ dependence of the dust sublimation radius, as understood in the `receding torus' concept \citep{law91,sim98,sim05,lus14}, without requiring any change in the scale height of the torus.
In contrast, the X-ray absorbed fraction remains nearly constant, falling only from 60\% to 45\%.

The implicit population of X-ray absorbed Sy\,1s were discussed by \cite{mer14} who argued that the X-ray absorption is not occuring on large galaxy-wide scales and suggested instead that it is more plausibly due to dust-free gas within the BLR.
Below, we argue that this is indeed likely, since BLR clouds are expected to contain a significant column of neutral gas; and associate the absorption with a `neutral torus' that co-exists with the BLR, and which has an impact on how the `receding torus' concept should be understood.

However, first it is important to clarify whether warm absorbers, which have been observed in about 50\% of Sy\,1s \citep{kom99,blu05,win12,lah14}, may be responsible for the X-ray absorbed Sy\,1s here.
Warm absorbers imprint their signature on an X-ray spectrum through absorption lines and are most prominent in soft X-ray bands.
They are classically identified by the O\,VII and O\,VIII absorption edges at 0.74\,keV and 0.87\,keV respectively, rather than through their impact on the spectral shape \citep{kom99,win12,lah14}. 
Warm absorbers have a smaller impact than neutral gas on the global shape of the spectrum, as evident from the models of \cite{pag11} whose fits to 0.2--10\,keV QSO spectra show that a warm absorber requires an order of magnitude more gas than a neutral absorber to reproduce the same spectral shape.
The median ionised column derived by \cite{win12} for the {\em Swift-BAT}, AGN as well as by \cite{lah14} on a different sample, is $\sim10^{21}$\,cm$^{-2}$; with columns an order of magnitude lower for AGN without strong O\,VII and O\,VIII detections.
While significant, this is less than $\sim10^{21.5}$\,cm$^{-2}$ threshold adopted here for defining X-ray absorbed systems.
Coupled with the fact that \cite{mer14} used either the full 0.5-10\,keV band or the 0.5--2\,keV versus 2--10\,keV hardness ratio to derive the absorbing column from its effect across the whole band rather than in specific features, 
this strongly argues that the absorptions they derived are due to neutral rather than ionised gas.

The BLR provides an ample source of neutral gas that could be responsible for the derived absorption.
Photoionisation models by \cite{net13} show that for a typical BLR ionisation parameter of $U \sim 0.01$ \citep{lei07,neg13} and a density of $10^{10}$\,cm$^{-3}$ the temperature drops below $10^4$\,K at a column of $\sim10^{21}$\,cm$^{-2}$ and the fractional abundance of H\,I increases dramatically.
Thus, for clouds with a column density of $10^{23}$\,cm$^{-2}$ as typically expected, the majority of gas is neutral rather than ionised.
This has been used to explain the cause of Sy\,1s which change state on short timescales, in terms of a BLR cloud passing across the line-of-sight to the central engine (see \citealt{ris10} and \citealt{tor14}, and references therein).
In this interpretation, the clouds have cores with column densities of at least a few $\times10^{23}$\,cm$^{-2}$ that cause the neutral absorption, and ionised tails that are the origin of the warm absorption \citep{ris09,ris11}.

Since the BLR clouds contain abundant neutral gas, it is natural to associate them with the X-ray absorbed Sy\,1s.
The remarkable constancy of the fraction of X-ray absorbed AGN across the luminosity range in Fig.~\ref{fig:syfrac} suggests that the geometry of the cold gas absorber also changes rather little: both the inner edge and scale height of the thick gas distribution remain similar to those at low luminosities.
The implication is that there is a `neutral torus' that co-exists with the BLR, and extends out to the start of the dusty molecular torus as illustrated in the righthand side of Fig.~\ref{fig:cartoon}.
In this context, we note that \cite{min15} and \cite{gan15} have independently suggested the core of the fluorescent Fe K$\alpha$ line may originate at radii extending from the BLR out to the dusty torus, implying the presence of gas at intermediate radii.
There is no clear boundary between the BLR and the neutral torus since it is the same clouds that contribute to both.
The BLR emission will follow a radial dependence associated with the ionisation parameter.
Since $U \propto L_{AGN}/(R^2 \ n_H)$, the rapid decrease of $U$ with radius $R$ may be largely mitigated if the cloud density $n_H$ also decreases with $R$.
Thus the BLR could extend far into the neutral torus.
How far it extends is given by the scaling of the BLR size with AGN luminosity \citep{kas00}.
Using the most recent measurement of that relation \citep{ben13}, and adopting a ratio $L_{AGN} / \lambda L_{5100} \sim 15$ \citep{gru04}, we estimate the characteristic radius of the BLR to be a factor of a few less than the dust sublimation radius.
However, it is also possible that the BLR emission could still occur as far as the dust sublimation radius.

In the context above, the idea of the `receding torus', in which the location of the inner wall of the torus is set by the dust sublimation radius, can be misleading.
We argue that the inner dust boundary and the inner gas boundary of the torus should be considered separately.
The discussion above suggests that the location of the inner gas boundary of the geometrically thick structure (causing the X-ray absorption), is roughly independent of AGN luminosity in the range $43 < \log{L_{14-195}} < 45$.
And it is within this geometrically thick gas structure that the inner dust boundary (i.e. the `receding torus') simply represents the location at which the predominant gas phase changes from ionised/neutral to dusty/molecular.
The thick gas structure itself remains nearly unchanged.
This means that at low luminosities, the thick gas structure is synonymous with the standard dusty torus;
but at high luminosities, the thick gas structure has an outer part that is the dusty torus, as well as an inner part that contributes to the BLR emission but also acts as a neutral dust-free torus.

Intriguingly, \cite{win12} found that the strength of the O\,VII and O\,VIII edges tracing the warm absorbers in {\em Swift-BAT} AGN are correlated with the neutral column density -- suggesting that material in the neutral torus is associated with a more-or-less proportional amount of ionised material.
This can be put in the context of AGN showing rapid variations in absorbing column, in which neutral clouds have an ionised outflowing tail \citep{ris09,ris11} -- an interpretation that applies to individual clouds in the BLR.
Given that warm absorbers are outflowing \citep{blu05,lah14}, and that the distance of the warm absorbers from the AGN is typically larger than the BLR but within a factor of a few of the dust sublimation radius \citep{blu05}, one might speculate that the ionised material seen by \cite{win12} is associated with the neutral torus and outflowing from the clouds as they are ablated.

\section{Conclusions}
\label{sec:conc}

We have described the rationale for defining a complete volume limited sample of bright local Seyferts, selected from the 14--195\,keV {\it Swift-BAT} catalogue.
These AGN are complemented by a matched sample of inactive galaxies, and we have shown that the host galaxy properties (stellar mass, morphological type, inclination, presence of a bar, distance) of these two samples exhibit similar distributions.
Spatially and spectrally resolved observations of the inner few hundred parsecs in these active and inactive galaxies will be used to study gas inflow and outflow and the processes regulating it.
The main points we have discussed here are:
\begin{itemize}

\item
A comparison of the properties of the Sy\,1s and Sy\,2s shows that there is no bias for type in the selection.
The only significant difference in AGN properties is the absorbing column $N_H$, the mean of which is two orders of magnitude greater for the Sy\,2s.
This is consistent with unification schemes in which optical obscuration is greater for Sy\,2s, but cannot distinguish between simple orientation dependence and instrinsic differences in torus properties.

\item
The fraction of Sy\,1s versus Sy\,2s in this sample, and also in the flux limited 14--195\,keV sample of \cite{win09}, demonstrates that the true fraction of Sy\,2s at $\log{L_{14-195}}\sim42.5$--44 is 50--60\%, consistent with \cite{law10} and the spectroscopic subsample of \cite{mer14}.
We show that the fraction of X-ray unabsorbed Sy\,2s is at most a few percent and hence that, at these luminosities, optical obscuration and X-ray absorption usually occur together.

\item
At higher luminosities $\log{L_{14-195}} > 44$, while the optically obscured fraction drops rapidly, the X-ray absorbed fraction remains the same.
We argue this implies that the inner boundary of the geometrically thick gas structure associated with the torus is roughly independent of AGN luminosity, having similar radial and height scales as for lower luminosities.
At low luminosities this gas structure is synonymous with the standard dusty torus;
but at high luminosities it has an outer part which is the dusty torus, and an inner part which is a neutral dust-free torus that also contributes to the broad line region emission.
In this context, the `receding torus' model simply represents the location within this gas structure at which the predominant phase changes from ionised/neutral to molecular/dusty.

\item
Finally, we note that the consistency of local (e.g. {\it Swift-BAT}) and more distant (e.g. the spectroscopically classified AGN in the $0.3<z<3.5$ sample of \citealt{mer14}) samples suggests that the obscuring structure does not strongly evolve with redshift.

\end{itemize}

\acknowledgments

The authors thank the referee for a variety of useful suggestions that have helped improve the paper.
They are also grateful to K.~Dodds-Eden for her major contribution to selecting the matched inactive sample.
R.D. thanks A.~Merloni and H.~Netzer for useful discussions.
E.K.S.H. acknowledges support from the NSF Astronomy and Astrophysics Research Grant under award AST-1008042.
C.R. acknowledges financial support from the CONICYT-Chile ``EMBIGGEN'' Anillo (grant ACT1101).
R.R. thanks CNPq for financial support.
M.K. acknowledges support from the Swiss National Science Foundation (SNSF) through the Ambizione fellowship grant PZ00P2\textunderscore154799/1.
This research has made use of the NASA/IPAC Extragalactic Database (NED) which is operated by the Jet Propulsion Laboratory, California Institute of Technology, under contract with the National Aeronautics and Space Administration.  
It also makes use of data products from the Two Micron All Sky Survey, which is a joint project of the University of Massachusetts and the Infrared Processing and Analysis Center/California Institute of Technology, funded by the National Aeronautics and Space Administration and the National Science Foundation.\\



\end{document}